\documentclass[preprint,tightenlines,amsmath,amssymb,showpacs,superscriptaddress,nofootinbib]{revtex4}
\usepackage{graphicx}
\usepackage{epstopdf}
\usepackage{dcolumn}
\usepackage{color}
\usepackage{bm}
\usepackage{overpic}
\usepackage{rotating}
\usepackage{times}
\usepackage{multirow}
\usepackage{float}
\usepackage{mathrsfs}
\usepackage{booktabs}
\usepackage[mathlines]{lineno}

\newcommand{\pp}{\pi^+\pi^-}
\newcommand{\kk}{K^+ K^-}

\newcommand{\EE}{e^+e^-}

\newcommand{\jpsi}{J/\psi}

\newcommand{\ppbar}{p\bar{p}}

\newcommand{\LLbar}{\Lambda\bar{\Lambda}}

\newcommand{\chisq}{\chi^{2}}

\def\Journal#1#2#3#4{{#1} {\bf #2}, #3 (#4)}

\def\EPJC{Eur. Phys. J. C}

\begin{document}
\normalsize
\parskip=5pt plus 1pt minus 1pt
\hyphenpenalty=10000
\tolerance=1000

\title{\boldmath Observation of a near-threshold enhancement in the $\Lambda\bar{\Lambda}$ mass spectrum from $e^+e^-\to\phi\Lambda\bar{\Lambda}$ at $\sqrt{s}$ from 3.51 to 4.60 GeV}

\author{
M.~Ablikim$^{1}$, M.~N.~Achasov$^{10,c}$, P.~Adlarson$^{67}$, S. ~Ahmed$^{15}$, M.~Albrecht$^{4}$, R.~Aliberti$^{28}$, A.~Amoroso$^{66A,66C}$, M.~R.~An$^{32}$, Q.~An$^{63,49}$, X.~H.~Bai$^{57}$
% ...
\\
\vspace{0.2cm}
(BESIII Collaboration)\\
\vspace{0.2cm} {\it
$^{1}$ Institute of High Energy Physics, Beijing 100049, People's Republic of China\\
% ...
$^{4}$ Bochum Ruhr-University, D-44780 Bochum, Germany\\
% ...
$^{10}$ G.I. Budker Institute of Nuclear Physics SB RAS (BINP), Novosibirsk 630090, Russia\\
% ...
$^{15}$ Helmholtz Institute Mainz, Staudinger Weg 18, D-55099 Mainz, Germany\\
% ...
$^{28}$ Johannes Gutenberg University of Mainz, Johann-Joachim-Becher-Weg 45, D-55099 Mainz, Germany\\
% ...
$^{32}$ Liaoning Normal University, Dalian 116029, People's Republic of China\\
% ...
$^{49}$ State Key Laboratory of Particle Detection and Electronics, Beijing 100049, Hefei 230026, People's Republic of China\\
% ...
$^{57}$ University of Jinan, Jinan 250022, People's Republic of China\\
% ...
$^{63}$ University of Science and Technology of China, Hefei 230026, People's Republic of China\\
% ...
$^{66}$ University of Turin and INFN, (A)University of Turin, I-10125, Turin, Italy; (B)University of Eastern Piedmont, I-15121, Alessandria, Italy; (C)INFN, I-10125, Turin, Italy\\
$^{67}$ Uppsala University, Box 516, SE-75120 Uppsala, Sweden\\
% ...
$^{c}$ Also at the Novosibirsk State University, Novosibirsk, 630090, Russia\\
% ...
}
}

\date{\today}

\begin{abstract}

The process $\EE \to \phi \Lambda \bar{\Lambda}$ is studied using
data samples collected with the BESIII detector at
the BEPCII collider at center-of-mass energies $\sqrt{s}$ ranging from $3.51$ to $4.60~{\rm GeV}$ . 
An enhancement is observed near the threshold of $\Lambda \bar{\Lambda}$. 
The lineshape of this enhancement is studied in different approaches, including fit with a Breit-Wigner function or a reversed exponential function.
The Breit-Wigner function has a mass of $(2262 \pm 4 \pm 28)~{\rm{MeV}}/c^{2}$ and a width of $(72 \pm
5 \pm 43)~\rm{MeV}$,
where the quoted uncertainties are statistical and systematic, respectively.
The rising rate of the reversed exponential function is measured as $33\pm11\pm6~\rm MeV/c^2$.
For the $\Lambda\bar{\Lambda}$ system, the $J^{PC}$ quantum numbers of $0^{-+}$ and $0^{++}$ are rejected, 
while other $J^{PC}$ hypotheses are possible, according to
the helicity angle study. The 
energy-dependent cross section
of the $\EE \to \phi \Lambda \bar{\Lambda}$ process is measured
for the first time in this energy region, and contributions
from excited $\psi$ states and vector charmonium-like
$Y$-states are investigated.

\end{abstract}

\maketitle

\section{Introduction}

Early in this century, a number of exotic states have been
discovered~\cite{reviewXYZ} in final states with a quarkonium
and one or two light hadrons, or with heavy-flavor mesons. Among
these states, there are vector states with $J^{PC}=1^{--}$ which
are usually called $Y$ states, such as the
$Y(4260)$~\cite{Aubert:2005rm},
$Y(4360)$~\cite{belle_y4660,babar_y4360}, and
$Y(4660)$~\cite{belle_y4660}. The $Y(4260)$ state is observed for
the first time by the BABAR experiment with a mass of
$(4259\pm8^{+2}_{-6}) {\rm~MeV}/c^2$ using the initial state
radiation (ISR) events $\EE\to\gamma_{\rm ISR} \pi^{+}\pi^{-}
J/\psi$~\cite{Aubert:2005rm}. The observation was latter confirmed
by the CLEO~\cite{cleoY4260} and Belle experiments~\cite{Yuan:2007sj}.
In 2017, a dedicated analysis performed by the BESIII experiment
revealed that the so-called $Y(4260)$ state is not simply one
Breit-Wigner (BW) resonance and can be a
combination of two states~\cite{Ablikim:2016qzw}. The first one
has a lower mass and a much narrower width than the $Y(4260)$,
but is consistent with the $Y(4220)$ state observed in $\EE\to
\pi^{+}\pi^{-} h_{c}$ events~\cite{Ablikim:2013wzq,czy}, and the second
one at around $4.32 {\rm~GeV}/c^2$ is observed for the first time
with a significance greater than $7.6\sigma$. The lower-mass
resonance was also observed in
$\EE\to\omega\chi_{c0}$~\cite{Ablikim:2019apl} and  $\pi \bar{D}
D^{*}+c.c.$ events~\cite{Ablikim:2018vxx}.

Until now, the $Y(4260)$ and other vector charmonium-like states
were only reported in final states containing a $c\bar{c}$ pair:
either charmonium states or charmed mesons. Several analyses
have been performed by the BESIII collaboration to search for
light hadron decays of these states, for example, $Y(4260)\to
\pi^0(\eta) p\bar{p}$~\cite{Ablikim:2017gtb}, $K^+ K^-
\pi^0$~\cite{Ablikim:2018jbb}, $\Xi\bar{\Xi}$~\cite{2019WXF}, {\it
etc.} Although there is no significant contribution from the
vector charmonium or charmonium-like states identified, the cross
section lineshapes of these processes do suggest contributions
from amplitudes beyond simple continuum production. In
Ref.~\cite{Maiani:2005pe}, the $Y(4260)$ state is interpreted as a
diquark-antidiquark state ([$cs$][$\bar{c}\bar{s}$]). This
interpretation implies that the $Y(4260)$ state decays easily into
final states containing a pair of $s\bar{s}$. 
One of the dominant contributions to $\EE\to \pi^{+}\pi^{-} J/\psi$ comes from $Y(4260)\to f_0(980)J/\psi$ decays ~\cite{zc3900_jpc},
and
the $f_0(980)$ meson is known to have a large $s\bar{s}$ component.
If the $c\bar{c}$ quarks in $Y(4260)$ annihilate while the $s\bar{s}$ pair
survives in the final state, we expect $Y(4260)$ decays into
strange mesons or baryons, such as $\phi\LLbar$. 
Such a signal will manifest itself as a lineshape distortion due to the interference
between the amplitudes of the $Y(4260)$ decay and the $\EE\to \phi\LLbar$ continuum production.

The $\eta(2225)$ and $\phi(2170)$ states~\cite{bes3white} are interpreted as loosely
bound states of $\LLbar$ in Ref.~\cite{Zhao:2013ffn}. This
suggests that the $\eta(2225)$ couples to $\LLbar$ strongly above
the threshold and it can be produced in $\EE\to \phi\LLbar$
processes. Together with the strong enhancement of the $\EE\to \LLbar$
production cross section close to threshold~\cite{bes3_LLb},
this can help establish if there is any connection between
these two hadron molecule candidates. On the other hand, near-threshold enhancements are observed in several processes involving baryon-anti-baryon pairs
such as $\jpsi\to \gamma\ppbar$~\cite{X1835}, $B\to
K\ppbar$~\cite{B2Kppbar}, and $B^0\to K\LLbar$~\cite{B2KLLbar}.
There are a few interpretations for these phenomena, including
states near threshold as found in a model by Nambu and
Jona-Lasinio~\cite{CNY}, $J^{PC}=0^{\pm+}$ isoscalar states
coupled to a pair of gluons~\cite{Rosner:2003bm}, and low-mass
enhancements favored by the fragmentation
process~\cite{Rosner:2003bm}. The isoscalar $C=+$ $\LLbar$
threshold enhancement can be searched for in
$\EE\to\phi\Lambda\bar{\Lambda}$ and its spin-parity can be
determined by studying the angular distribution if the data sample is large enough.

In this paper, we report the first observation of the process
$\EE\to\phi\Lambda\bar{\Lambda}$ analysing data samples taken at
center-of-mass (CM) energies $\sqrt{s}$ ranging from 3.51 to 4.60~GeV. The
vector charmonium-(like) states are studied based on the energy-dependent cross sections, and an intermediate state in the
$\LLbar$ system is investigated to extract information on light
mesons.

\section{Detector and data samples}

The BESIII detector is a magnetic
spectrometer~\cite{Ablikim:2009aa} located at the Beijing Electron
Positron Collider (BEPCII)~\cite{Yu:IPAC2016-TUYA01}. The
cylindrical core of the BESIII detector consists of a helium-based
multilayer drift chamber (MDC), a plastic scintillator
time-of-flight system (TOF), and a CsI(Tl) electromagnetic
calorimeter (EMC), which are all enclosed in a superconducting
solenoidal magnet providing a 1.0~T magnetic field. The solenoid
is supported by an octagonal flux-return yoke with resistive plate
counter muon identifier modules interleaved with steel. The
acceptance of charged particles and photons is 93\% over $4\pi$
solid angle. The charged-particle momentum resolution at $1~{\rm
GeV}/c$ is $0.5\%$, and the $dE/dx$ resolution is $6\%$ for the
electrons from Bhabha scattering. The EMC measures photon energies
with a resolution of $2.5\%$ ($5\%$) at $1$~GeV in the barrel (end
cap) region. The time resolution of the TOF barrel part is 68~ps,
while that of the end cap part is 110~ps. The end cap TOF system
is upgraded in 2015 with multi-gap resistive plate chamber
technology, providing a time resolution of 60~ps~\cite{etof}.

The experimental data used in this analysis were taken at the CM
energies ranging from $3.51$ to $4.60$~GeV as shown in
Table~\ref{tab:CS}. Simulated samples produced with the {\sc
geant4}-based~\cite{geant4} Monte Carlo (MC) package which
includes the geometric description of the BESIII detector and the
detector response, are used to determine the detection efficiency
and to estimate the background contributions. The simulation includes the beam
energy spread and ISR in the $e^+e^-$
annihilations modeled with the generator {\sc kkmc}~\cite{kkmc}.
Inclusive MC simulation samples 
generated at $\sqrt{s}=4.178$ GeV are used to analyze the possible
background contributions. In total, these samples are 40 times larger than the data sample. They  
consist of open charm
production processes, ISR production of vector
charmonium(-like) states, and continuum processes ($\EE\to
q\bar{q}~,q=u,~d,~s$). The open charm production processes are
generated using {\sc conexc}, and the ISR production is
incorporated in {\sc kkmc}~\cite{kkmc}. The known decay states are
modeled with {\sc evtgen}~\cite{evtgen} using branching fractions
taken from the Particle Data Group (PDG)~\cite{PDG}, and the
remaining unknown decays from the charmonium states with {\sc
lundcharm}~\cite{lundcharm}. The final state radiations (FSR) from
charged final state particles is incorporated with the {\sc
photos} package~\cite{photos}.
It should be pointed out that only $\EE\to\kk\LLbar$ is simulated in the inclusive MC samples.

The signal MC samples of $\EE\to\phi\LLbar$ are generated 
with {\sc evtgen}~\cite{evtgen} along with {\sc kkmc}~\cite{kkmc} to handle the  $e^+e^-$ annihilations and ISR production.
The signal events are generated with three-body phase space (PHSP) 
model where the $\phi\LLbar$  distributed uniformly in the phase space.
The data samples used in this analysis have been collected by BESIII at 28 CM energies between
3.51 and 4.60~GeV, as listed in Table~\ref{tab:CS}, along with the CM energy and corresponding integrated luminosity.
The total integrated luminosity is $19.5~\rm fb^{-1}$.

\section{Event Selection}

The selection of charged tracks is based on the following
criteria. For each charged track, the polar angle in the
MDC must satisfy $|\cos\theta|<0.93$, and the
point of closest approach to the $\EE$ interaction point (IP) must be
within $\pm 20$~cm in the beam direction and within $10$~cm in the
plane perpendicular to the beam direction.
The particle identification (PID) of kaons, pions, and protons is based on the $dE/dx$ and TOF information.
Assumption of a given particle identification is based on the largest of the all PID hypotheses probabilities.
The $\phi$ meson is reconstructed using candidate $\kk$ pairs.
One $\bar{\Lambda}$ ($\Lambda$) baryon is assumed to be missing in order to improve the reconstruction efficiency.
Thus we require that there should be at least one proton and one pion with opposite charge, and one $\kk$ pair in the final state.

Since the $\Lambda$ baryon has a relatively long lifetime, it travels a certain distance before it decays.
A vertex fit is applied to its decay products $p\pi^-$ ($\bar{p}\pi^+$)
to ensure that their tracks are pointing back to the same vertex.
The $\Lambda$ ($\bar{\Lambda}$) baryon is reconstructed combining the $p\pi^-$ ($\bar{p}\pi^+$) final state passed the vertex fit.
Then, to verify that the selected $\kk\Lambda(\bar{\Lambda})$ candidates originate from the IP, another vertex fit is performed.
Only events with a good quality vertex fit are retained.
The flight distance between the IP and the $\Lambda$ decay vertex is required to be greater than two times of its resolution.
The momenta corrected by the vertex fit are used for kinematic fit.

\begin{figure*}[htbp]
  \begin{center}
  \centering
  \includegraphics[width=0.33\textwidth]{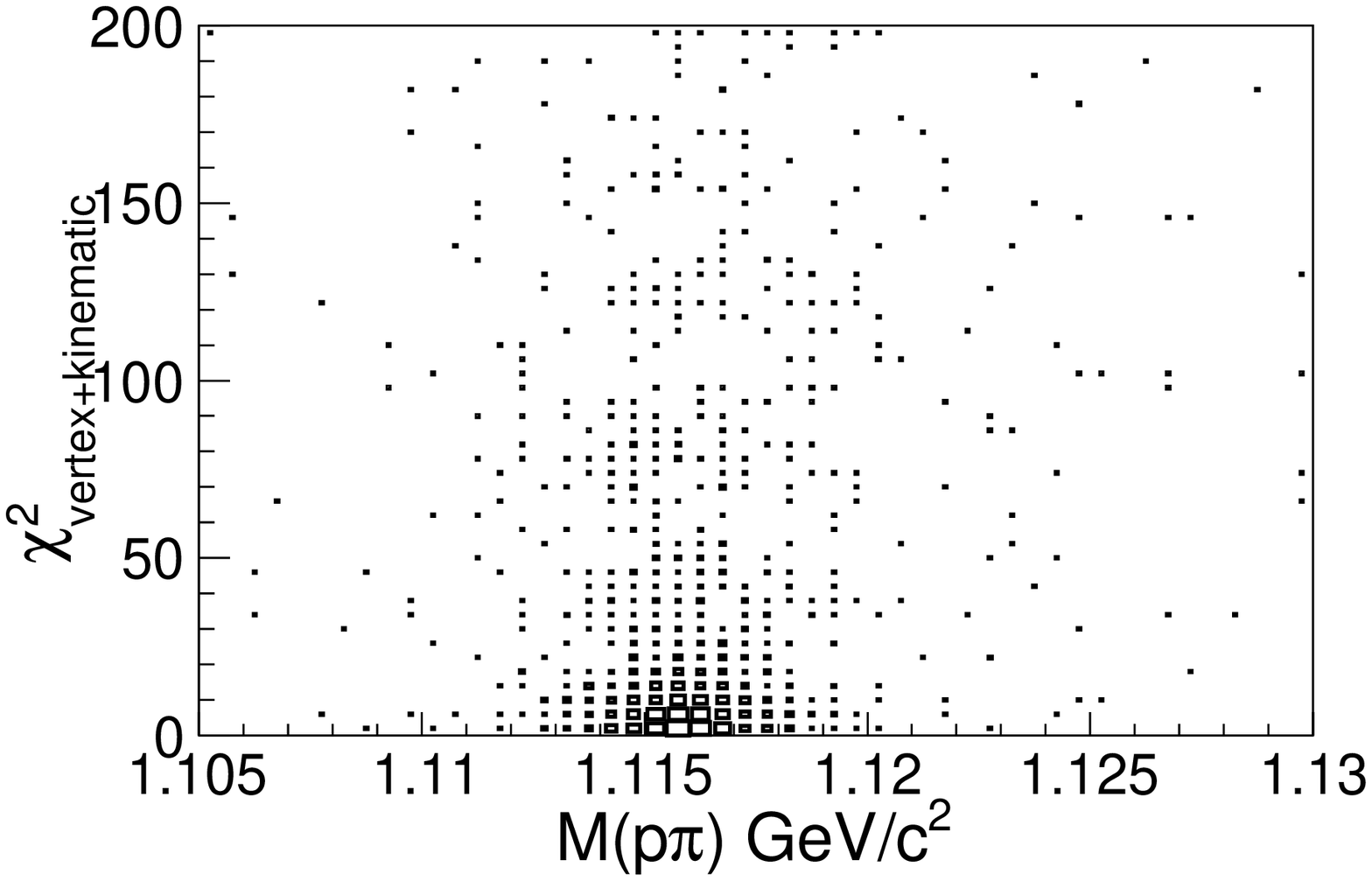}
  \includegraphics[width=0.33\textwidth]{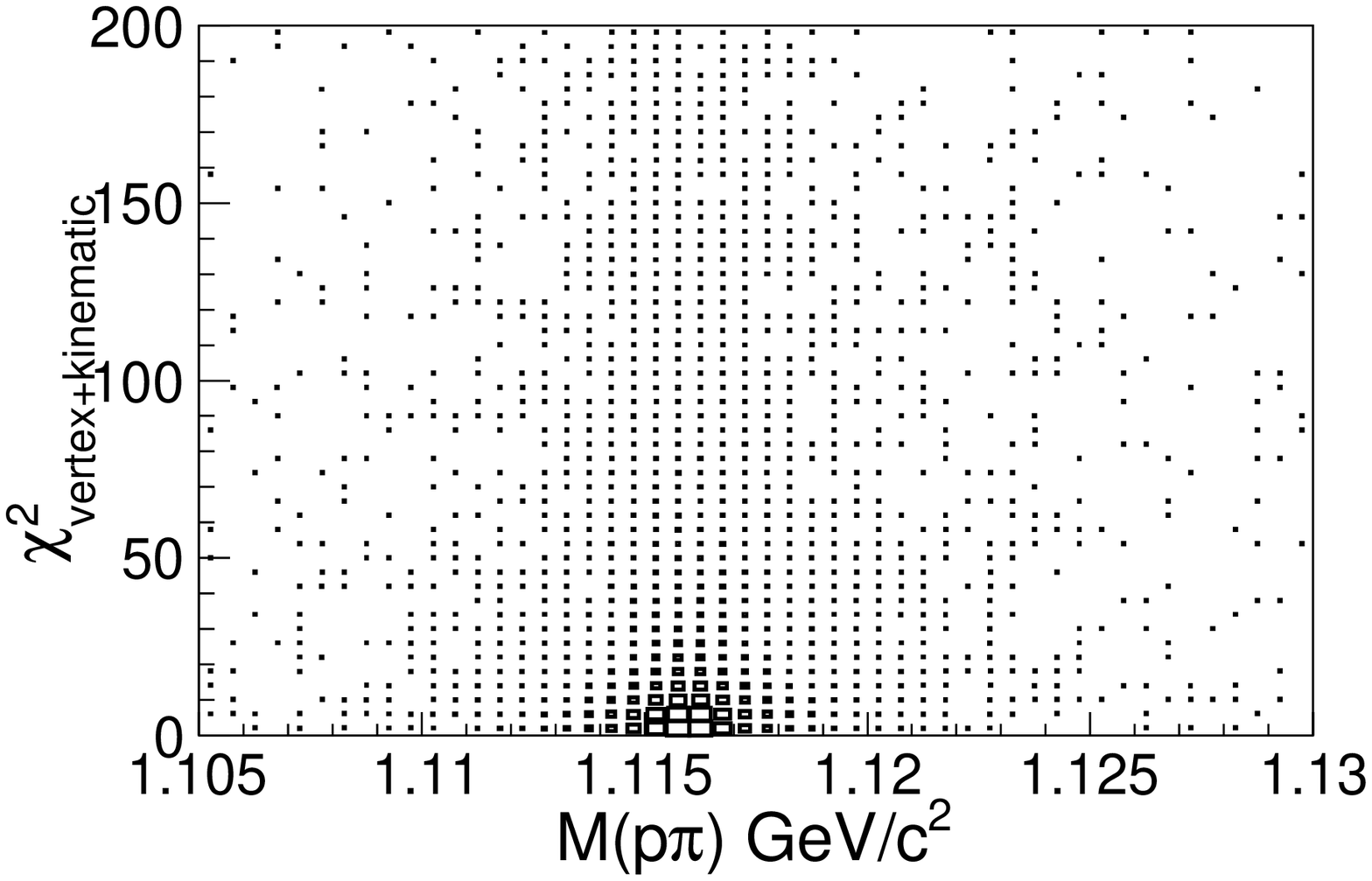}
  \includegraphics[width=0.33\textwidth]{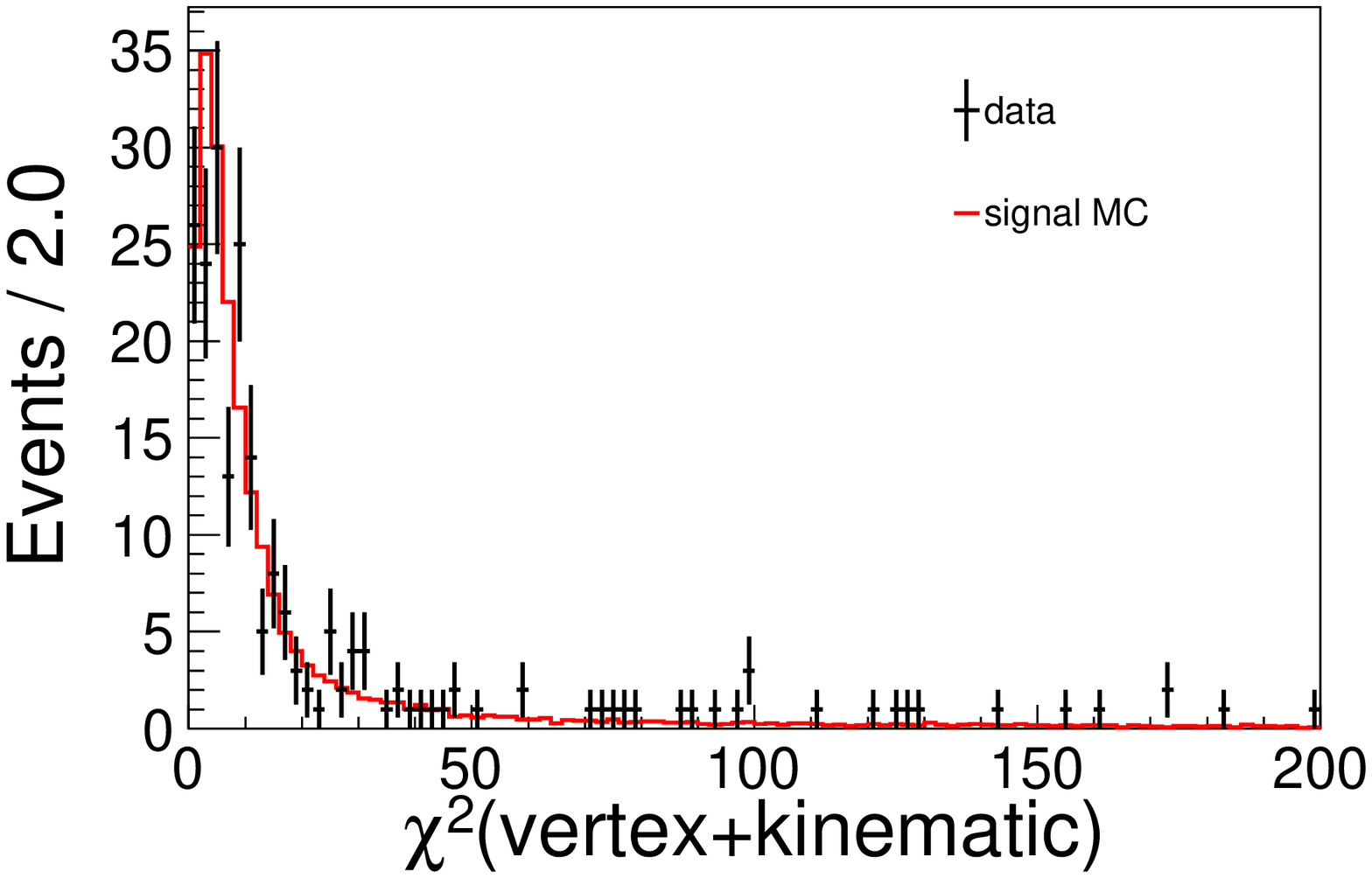}
  \caption{Scattering plots of $\chi^2$ versus the invariant mass of $p\pi$ from data sample summing all energy points (left) and MC simulation at 4.178 GeV (middle). Right plot is comparison of the $\chi^2$ distribution between data and MC samples.}
  \label{chi2MLambda}
  \end{center}
\end{figure*}

To improve the track momentum resolution and to reduce the background, 
a kinematic fit is applied
to the $\kk\Lambda(\bar{\Lambda})$ candidates constraining the missing mass
to the nominal mass of $\Lambda$.
The fraction of events containing more than one $\Lambda$ baryon is about 31\%.
For events with multiple candidates,
we choose the combination with the smallest $\chi^2$ combining the two vertex fits, and the kinematic fit.
The distributions of the combined $\chi^2$ versus the invariant mass of $p\pi$ are shown in Fig.~\ref{chi2MLambda}.
To make a better comparison, the one dimensional $\chi^2$ distribution compared between data and MC is shown in Fig.~\ref{chi2MLambda} (right). The possible difference are considered as systematic uncertainty coming from the kinematic fit.
The sum of the $\chisq$ values of the vertex and kinematic fits is required to be less than 30.
The invariant mass of selected $p\pi$ final state should be within the interval $[1.112,~1.120]~\rm GeV/c^2$, which covers about $\pm3\sigma$ of the $\Lambda$ signal region.

\begin{table*}[!htb]
  \centering
  \caption{Summary of the cross section measurements of the $\EE \to \phi \Lambda \bar{\Lambda}$ process at each CM energy point. Here, $\mathcal{L}_{\rm{int}}$ is the integrated luminosity, $N_{\rm sig}$ is the number of signal events from the fit to $M(K^+K^-)$ distributions with statistical uncertainty only, $\varepsilon$ is the efficiency, $(1+\delta)$ is the radiative correction, and $\sigma^{\rm dress}~(\rm pb^{-1})$ is the cross section quoted with a statistical and systematic uncertainty, respectively.}
  \label{tab:CS}
  \begin{tabular}{ c c c c c c c}
  \hline\hline
$\sqrt{s}$ (MeV)  &  $\mathcal{L}_{\rm{int}}~(\rm pb^{-1})$  &  $N_{\rm{\rm sig}}$  &  $\varepsilon$~(\%) &  $(1+\delta)$ &  $\sigma~(\rm pb)$ \\ \hline

3510.6  &  366.1 & $4.28\pm2.19$  &  4.22 &  0.84 &  $0.66\pm0.34\pm0.05$ \\
3773.0  &  2931.8 & $167.79\pm14.29$  &  10.93 &  0.89 &  $1.20\pm0.10\pm0.11$ \\
3869.5  &  224.0 & $15.02\pm3.69$  &  12.27 &  0.91 &  $1.23\pm0.31\pm0.11$ \\
4007.6  &  482.0 & $34.83\pm6.28$  &  13.90 &  0.95 &  $1.14\pm0.20\pm0.09$ \\
4128.5  &  401.5 & $17.73\pm4.49$  &  13.55 &  1.06 &  $0.64\pm0.16\pm0.05$ \\
4157.4  &  408.7 & $19.01\pm4.95$  &  14.22 &  1.02 &  $0.69\pm0.17\pm0.06$ \\
4178.4  &  3160.0 & $173.23\pm14.34$  &  14.63 &  1.00 &  $0.76\pm0.07\pm0.07$ \\
4188.8  &  565.8 & $29.84\pm5.96$  &  14.76 &  1.00 &  $0.74\pm0.16\pm0.07$ \\
4198.9  &  524.6 & $23.57\pm5.38$  &  14.93 &  1.01 &  $0.62\pm0.13\pm0.05$ \\
4209.2  &  573.0 & $26.61\pm5.62$  &  14.47 &  1.01 &  $0.65\pm0.15\pm0.06$ \\
4218.7  &  568.9 & $28.75\pm5.97$  &  14.50 &  1.01 &  $0.64\pm0.14\pm0.06$ \\
4226.3  &  1100.9 & $66.49\pm8.89$  &  15.62 &  1.00 &  $0.77\pm0.11\pm0.07$ \\
4235.7  &  530.6 & $23.10\pm5.23$  &  15.01 &  1.03 &  $0.58\pm0.15\pm0.05$ \\
4243.8  &  537.4 & $15.88\pm4.53$  &  14.31 &  1.15 &  $0.36\pm0.11\pm0.03$ \\
4258.0  &  828.4 & $54.94\pm8.17$  &  14.56 &  1.11 &  $0.82\pm0.14\pm0.07$ \\
4266.8  &  529.7 & $29.22\pm6.06$  &  14.62 &  1.10 &  $0.71\pm0.14\pm0.06$ \\
4277.7  &  175.5 & $2.31\pm2.07$  &  14.15 &  1.09 &  $0.18\pm0.15\pm0.02$ \\
4287.9  &  502.4 & $18.48\pm4.70$  &  14.47 &  1.09 &  $0.47\pm0.12\pm0.04$ \\
4312.0  &  501.2 & $25.19\pm5.75$  &  15.54 &  0.97 &  $0.69\pm0.16\pm0.07$ \\
4337.4  &  505.8 & $25.26\pm5.55$  &  16.41 &  0.97 &  $0.62\pm0.14\pm0.06$ \\
4358.3  &  543.9 & $36.31\pm5.80$  &  17.09 &  0.96 &  $0.83\pm0.14\pm0.07$ \\
4377.4  &  522.7 & $28.33\pm5.79$  &  17.08 &  0.96 &  $0.70\pm0.15\pm0.06$ \\
4396.5  &  507.8 & $35.20\pm6.32$  &  16.89 &  0.97 &  $0.88\pm0.16\pm0.08$ \\
4415.6  &  1090.7 & $55.73\pm7.93$  &  17.44 &  0.96 &  $0.59\pm0.09\pm0.05$ \\
4436.2  &  569.9 & $35.80\pm6.74$  &  17.41 &  0.95 &  $0.66\pm0.13\pm0.06$ \\
4467.1  &  111.1 & $8.31\pm2.91$  &  17.80 &  0.95 &  $0.91\pm0.30\pm0.08$ \\
4527.1  &  112.1 & $7.98\pm2.81$  &  18.09 &  0.96 &  $0.80\pm0.28\pm0.07$ \\
4599.5  &  586.9 & $35.10\pm6.41$  &  16.30 &  0.97 &  $0.73\pm0.14\pm0.06$ \\

\hline\hline
  \end{tabular}
\end{table*}

\section{Data analysis}
\subsection{Signal extraction}

Studies of the inclusive MC simulation indicate that the main background contribution comes from the process
$\EE \to (\gamma) K^{+} K^{-} \Lambda \bar{\Lambda}$,
which does not peak around the $\phi$ signal area.
It should be pointed out that at $\sqrt{s}=4.6$~GeV, the CM energy is above the threshold of $\EE\to\Lambda^+_c\bar{\Lambda}^-_c$, and there is a background contribution from $\Lambda^{\pm}_c\to\Lambda K^{\pm}$ decays.
However, according to the Born cross section reported in Ref.~\cite{LcLc}, the contribution of this background in the whole fitting range (the invariant mass of $\kk$ $M(\kk)\in[0.98,~1.20]~\rm GeV/c^2$) is estimated to be only $8.8\pm0.1$ events.
Other sources of background are considered are found to be negligible. 

\begin{figure}[htbp]
\begin{center}
 \centering
 \includegraphics[width=0.45\textwidth]{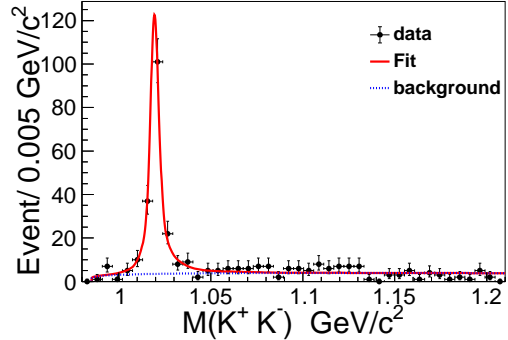}
 \caption{Invariant mass distribution of the $\phi\to\kk$ candidates for the data sample collected at $\sqrt{s}=4.178$ GeV. The data (dots) are overlaid by the result of the fit (red solid line) described in the text. The blue dotted line represents the background component of the fit.
 }
 \label{fig:FitPhi}
 \end{center}
\end{figure}

To obtain the signal yields, an unbinned maximum likelihood fit
is performed to the invariant mass spectrum of the $\kk$ pair for each CM energy point.
The signal distribution is described by a MC-simulated shape, and the background shape is described by an inverted
ARGUS~\cite{ARGUS} function whose threshold is fixed to $2m_{K^{\pm}}$, where $m_{K^{\pm}}$ is the nominal kaon mass~\cite{PDG}.
The fit result for $\sqrt{s}=4.178$~GeV is shown in Fig.~\ref{fig:FitPhi} as an example,
and the numbers of signal events ($N_{\rm sig}$) at 28 energy points are listed in Table~\ref{tab:CS}.

\subsection{Intermediate structure study}

\begin{figure}[htbp]
\begin{center}
 \centering
   \includegraphics[width=0.45\textwidth]{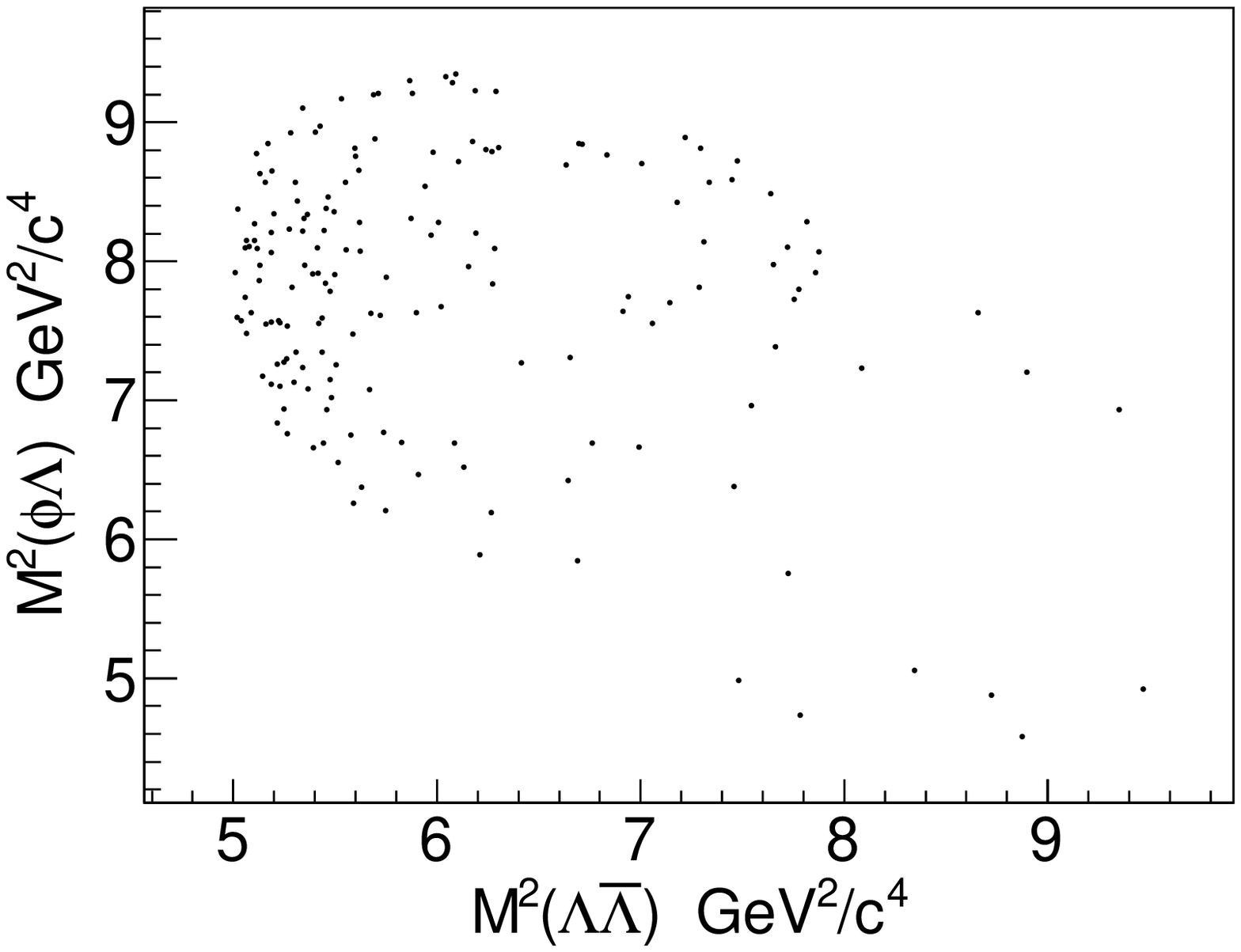}
 \caption{Dalitz plots for the $\EE\to\phi\LLbar$ events
 at $\sqrt{s}=4.178~\rm GeV$. }
 \label{fig:Dalitz_data}
 \end{center}
\end{figure}

\begin{figure}[htbp]
\begin{center}
 \centering
 \includegraphics[width=0.45\textwidth]{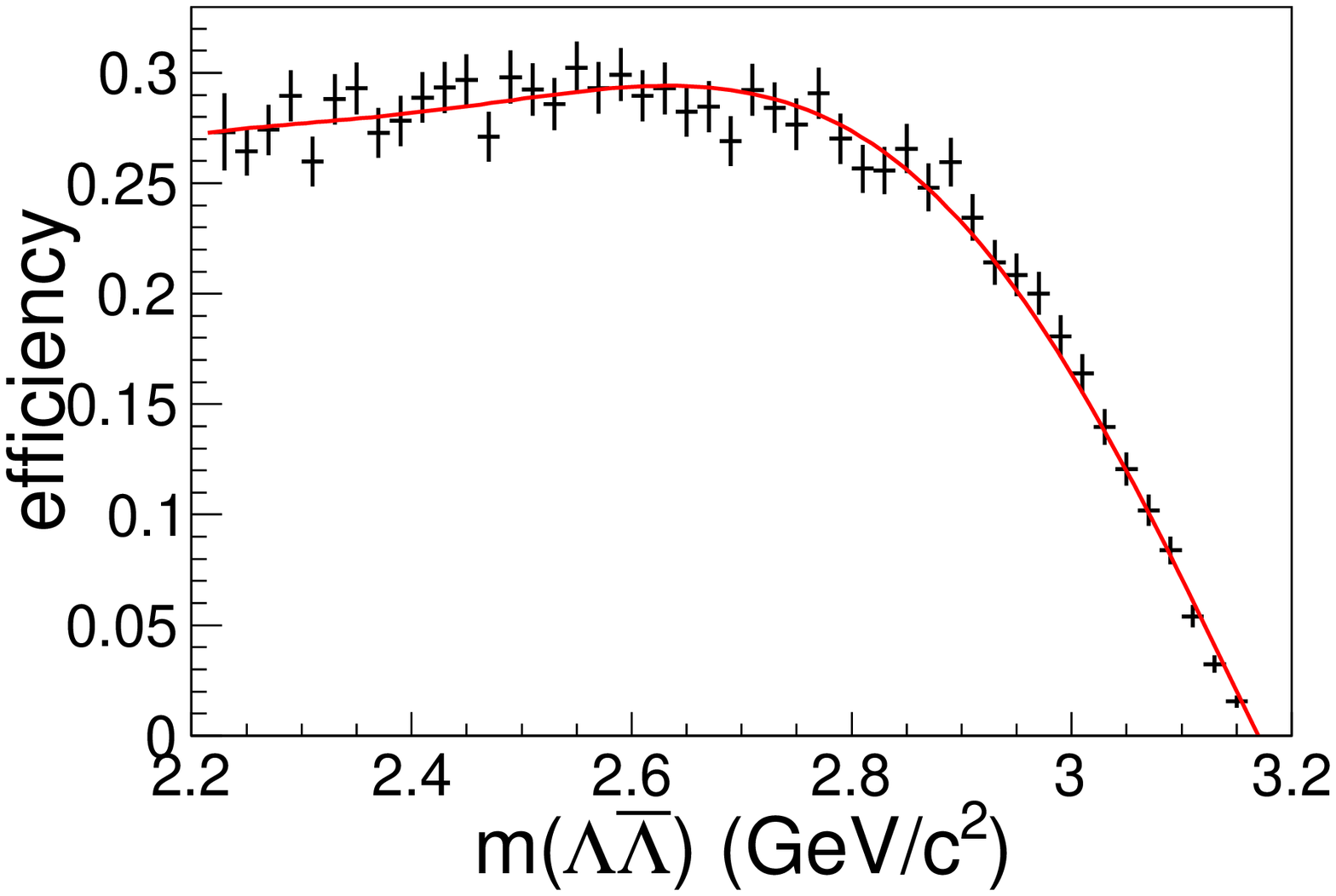}
 \includegraphics[width=0.45\textwidth]{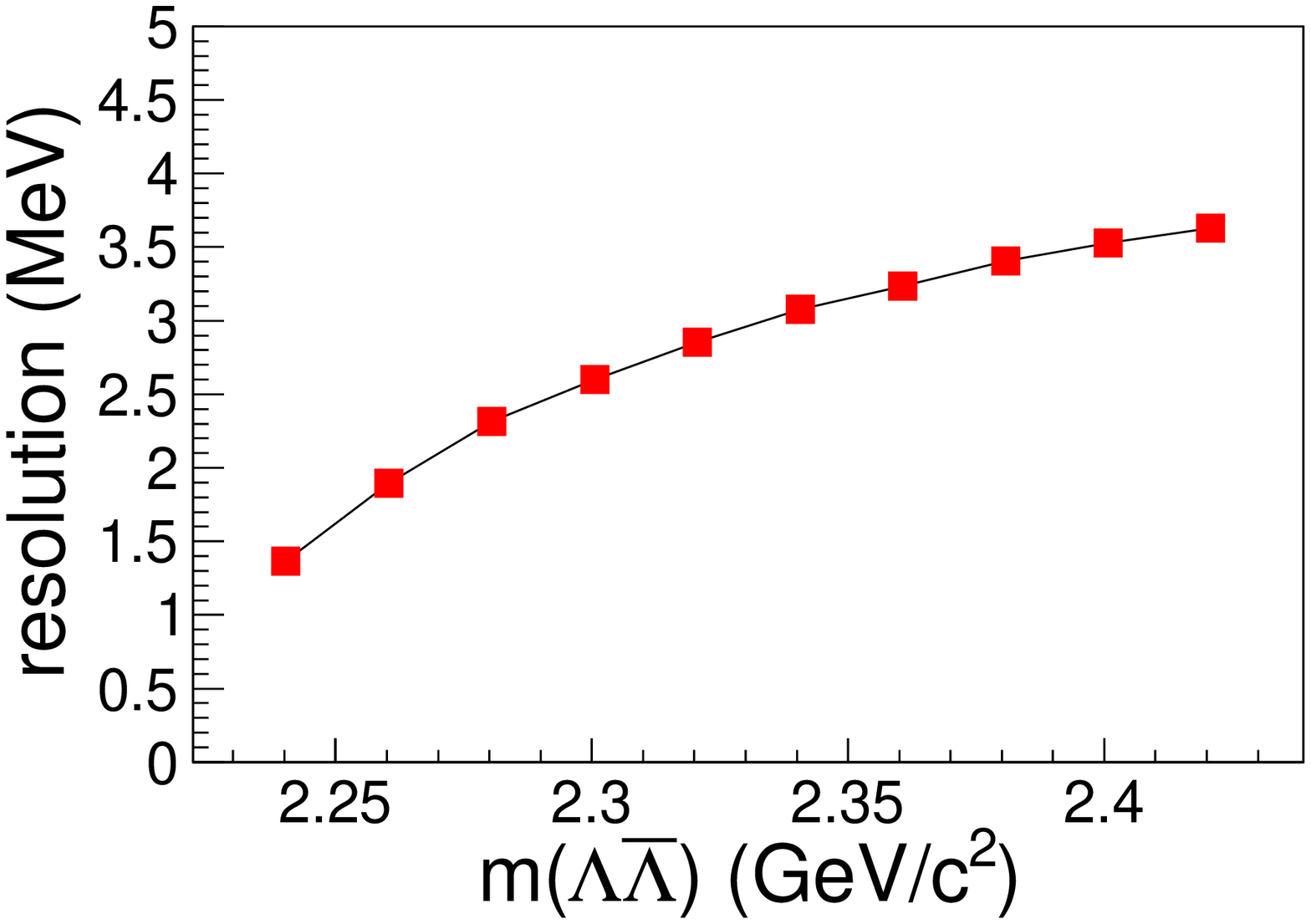}
 \caption{Mass-dependent reconstruction efficiency (top) and resolution (bottom) curve from the signal MC study at $\sqrt{s}=4.178~\rm GeV$.}
 \label{fig:eff_res_MLL}
 \end{center}
\end{figure}

We perform a study to investigate possible intermediate structures to better estimate the reconstruction efficiency.
The Dalitz plot distribution of the $\phi\LLbar$ candidates at $\sqrt{s}=4.178$ GeV is shown in Fig.~\ref{fig:Dalitz_data}, after requiring that $M(\kk)\in[1.01,~1.03]~\rm GeV/c^2$.
It is clear that most of the events in the data are deposited near the $\LLbar$ threshold, which is different from the PHSP MC sample generated with a uniform distribution.
Signal MC samples are generated at 28 energy points to study the reconstruction efficiency and resolutions. The efficiency and resolution curves are shown in Fig.~\ref{fig:eff_res_MLL} for MC samples at $\sqrt{s}=4.178~\rm GeV$.
We can see that the reconstruction efficiency is quite smooth near the threshold and the resolution is relatively small.

\begin{figure*}[htbp]
\begin{center}
 \centering
         \includegraphics[width=0.45\textwidth]{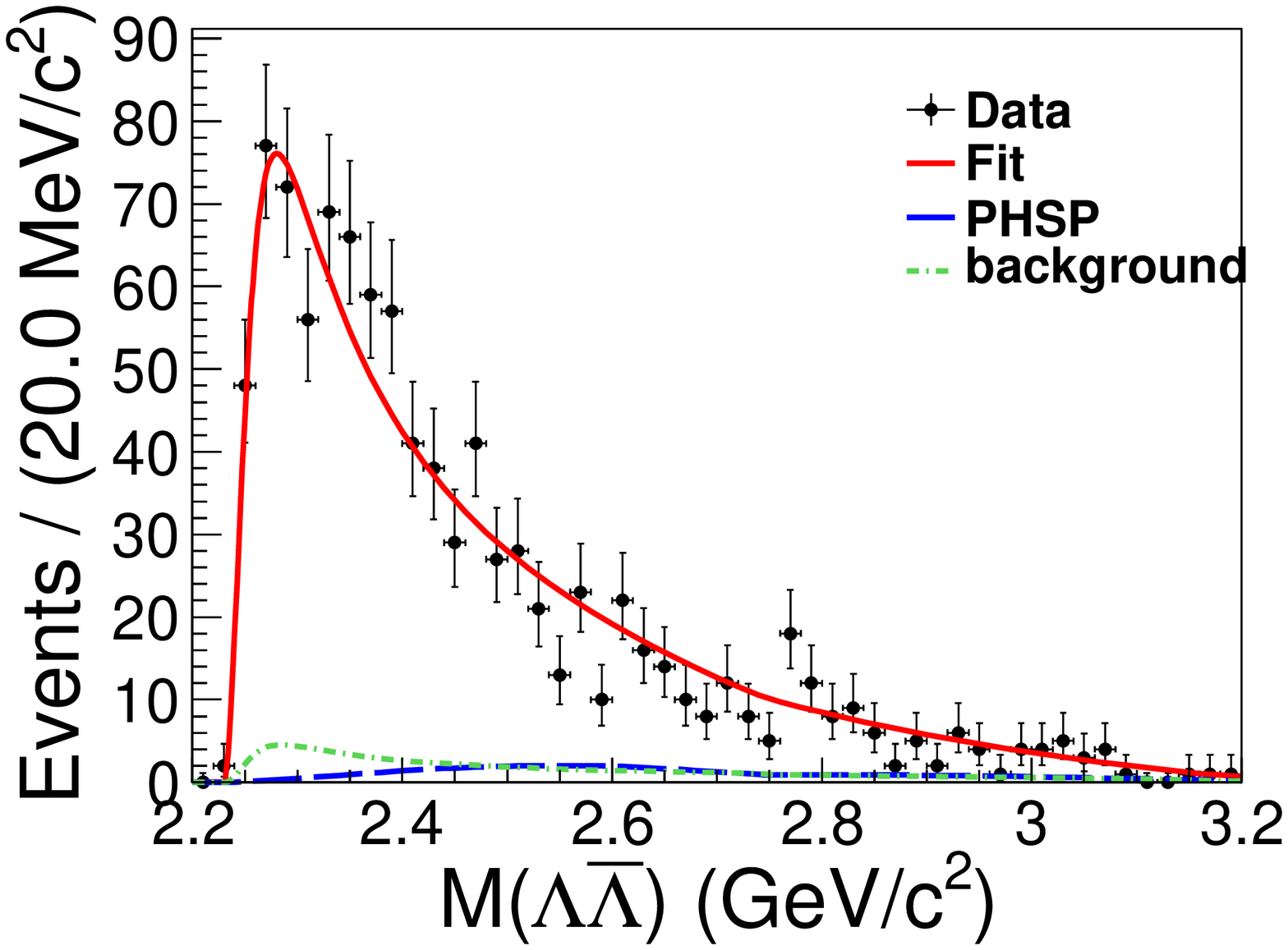}
         \includegraphics[width=0.45\textwidth]{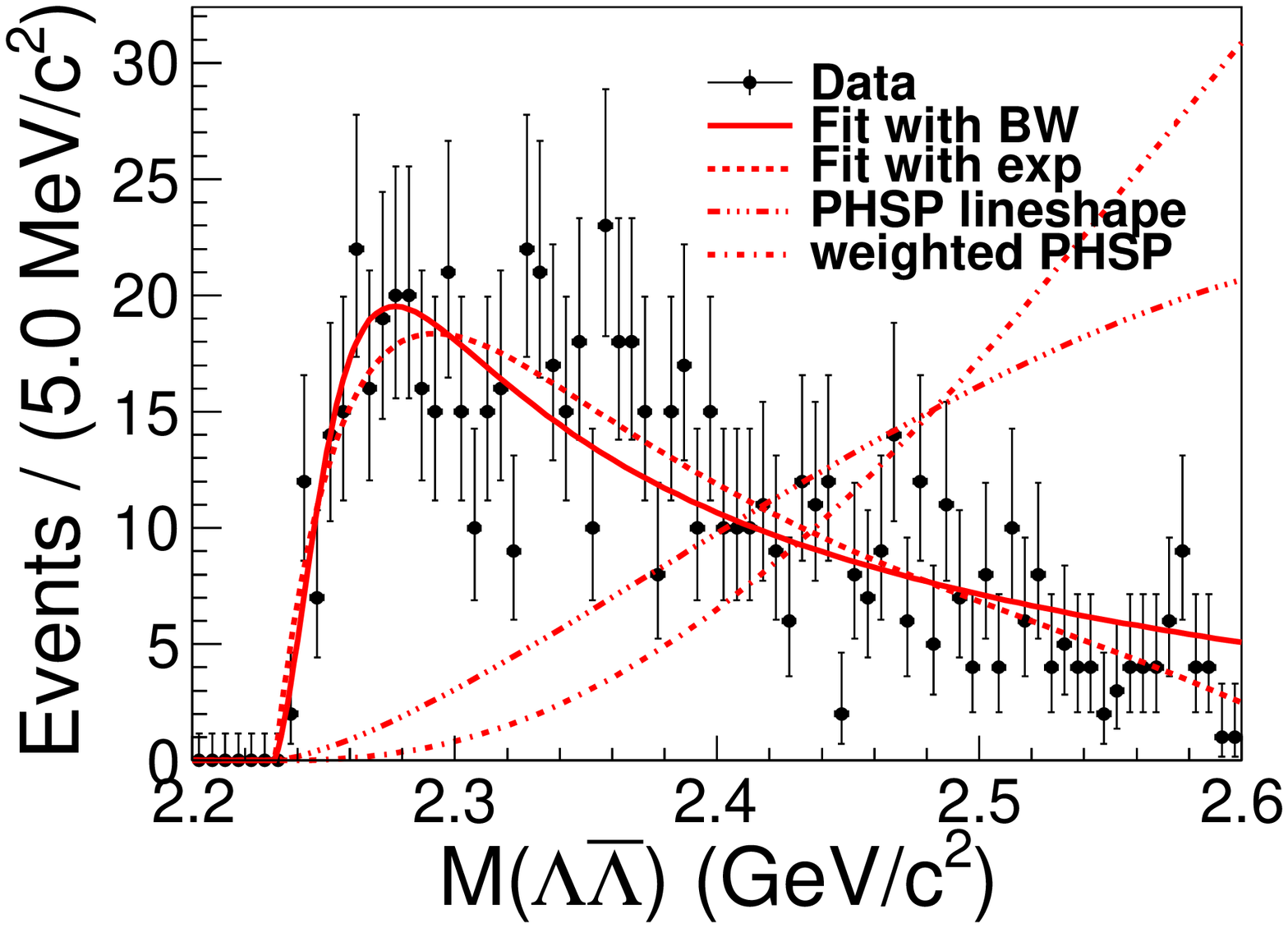}
 \caption{Invariant mass distribution of the $\LLbar$ candidates for the full data sample. The data (dots) are overlaid by the result of the fit (red solid line) described in the text. The blue long dashed line represents the PHSP components, and the green dashed-dotted line represents the background. The same distribution focusing on the lower mass range is shown in the right plot, where the red dashed line represents the fit result with the reversed exponential function, the red dash triple dotted line represents the PHSP lineshape, and red dash-dotted line represents the lineshape of PHSP events weighted by the angular distribution.
 }
 \label{fig:SimFit}
 \end{center}
\end{figure*}

The invariant mass distribution of the $\LLbar$ candidates for the full data sample is shown in Fig.~\ref{fig:SimFit}.
There are serval dynamics to generate such an enhancement, including final state interaction (FSI), a tail of a lower mass resonance, and so on~\cite{X1835,Rosner:2003bm,PDG}.
To describe the lineshape of this enhancement, an extended unbinned maximum likelihood fit is performed to all the data samples simultaneously.
We firstly perform a fit using a Breit-Wigner function (BW) to describe the signal.
Three components are considered in the fit: a near-threshold enhancement, a component distributed uniformly in PHSP, and a non-$\phi$ background component.
The interference between the resonant signal and non-resonant signal is ignored here.
The following formula is used to describe the lineshape of the enhancement~\cite{XXP:2014}:
\begin{equation}
\begin{aligned}
        & dN/dm_{\LLbar} \propto  \\
        &  \varepsilon(M_{\LLbar}) (k^{*})^{2l+1} f_{l}^{2}(k^{*}) |BW(M_{\LLbar})|^2 (q^{*})^{2L_{d}+1} f^{2}_{L_{d}}(q^{*})
\end{aligned}
\end{equation}
where
$\varepsilon(M_{\LLbar})$ is the mass-dependent efficiency obtained from MC simulation.
Here the MC sample is generated with the nonuniform angular distributions measured in data (to be described latter).$M_{\LLbar}$ is the invariant mass of $\Lambda\bar{\Lambda}$ system,
\begin{equation}
k^{*} \equiv {\sqrt{ \bigg(\frac{s+M^2_{\LLbar} - m^2_{\phi}}{2\sqrt{s}}\bigg)^2 - M^2_{\LLbar}}}
\end{equation}
is the momentum of $\LLbar$ system in the $\EE$ rest frame, where $m_{\phi}$ is the nominal mass of $\phi$~\cite{PDG},
\begin{equation}
\begin{aligned}
q^{*} \equiv {\sqrt{M^2_{\LLbar}/4-m^2(\Lambda)}}
\end{aligned}
\end{equation}
is the momentum of the $\Lambda$ baryon in $\LLbar$ system rest frame,
$l$ is the orbital angular momentum between $\phi$ and $\LLbar$ system, $L_{d}$ is the orbital angular momentum between $\Lambda$ and $\bar{\Lambda}$,
$f_{L}$ is the Blatt-Weisskopf barrier factor, with $f_{0}^{2}(z)=1$, $f_{1}^{2}(z)=1/(1+z)$, and $f_{2}^{2}(z)=1/(9+3z+z^2)$.
The relativistic Breit-Wigner function with a mass-dependent width used here is defined as
\begin{equation}
BW(M_{\LLbar}) \propto \frac{1}{M_{\LLbar}^2-m^2-im\Gamma_X},
\end{equation}
where $\Gamma_X\equiv\Gamma_0(q^{*}/q^{0})^{2L_{d}+1}(m/M_{\LLbar}) (f^{2}_{L_{d}}(q^{*})/f^{2}_{L_{d}}(q^{0}))^{2}$, $m$ and $\Gamma_0$ are the mass and width of the BW function, respectively, and $q^{0}$ is equal to  $q^{*}$ for $M(\LLbar)=m$.
In the fit, the mass and width are shared parameters between all the data samples, and are left free, as well as the signal yields.
The orbital angular momentum between $\phi$ and the $\LLbar$ system is $l=0$, and the orbital angular momentum between the $\Lambda$ and $\bar{\Lambda}$ baryons is $L_{d}=1$, assuming this is a $1^{++}$ or $2^{++}$ state.
Please note that even we use a BW function here to describe the near-threhold enhancement, we are not suggesting that this enhancement is a resonant-(like) structure.
The resolution effect is ignored here because it is relatively small compared with such a broader distribution.

The shape for PHSP signal is obtained from MC simulation. 
The shape of the non-$\phi$ background is obtained from the $\phi$ sideband region ($M(\kk)\in[0.99,~1.005]$ or $[1.075,~1.090]~\rm GeV/c^2$), and is parameterized with a Landau function. The number of background events is extrapolated from the sidebands to the $\phi$ signal region using the inverted ARGUS background function.
The fit result using all data samples~\cite{fitInSupply} is show in Fig.~\ref{fig:SimFit} (left). We also zoom in on the lower mass side to have a closer look on the rise of the enhancement, as shown in Fig.~\ref{fig:SimFit} (right).
The mass and width of the BW formula are fitted as $(2262\pm4)~{\rm MeV}/c^{2}$ and $(72\pm5)~\rm MeV$, respectively.

Alternatively, we perform a fit to estimate the rise rate near the threshold with the formula:
\begin{equation}
dN/dm_{\LLbar} \propto \mathcal{P}^{3}(1-e^{-\Delta M_{\LLbar}/p_{0}}),
\end{equation}
where $\mathcal{P}^{3}$ is a third-order polynomial whose parameters are free, $p_{0}$ is a free parameter, and $\Delta M_{\LLbar}\equiv M_{\LLbar}-2m_{\Lambda}$.
The fit result is shown in Fig.~\ref{fig:SimFit} (right), with $p_{0}=33\pm11~\rm MeV/c^2$.
Compared with the lineshapes of PHSP events weighted by angular distribution and cross section from each energy point, the rising rate in data is much faster.

\begin{figure*}[htbp]
\begin{center}
    \includegraphics[width=0.99\textwidth]{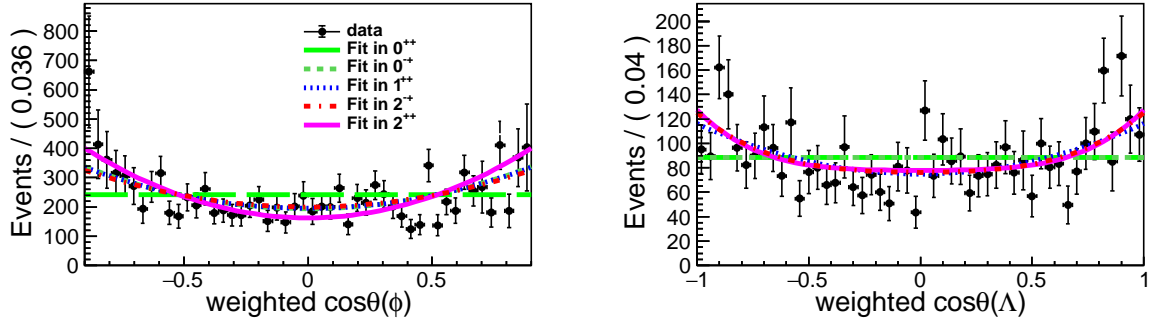}
\end{center}
\caption{Efficiency-corrected angular distribution of the $\phi$ (left) and $\Lambda$ (right) candidates combining all the data samples and the simultaneous fit results with different hypotheses of $J^{PC}$. 
The data (dots) are overlaid by the result of the fits described in the text, the green long dashed curves are fit result for $J^{PC}=0^{++}$, the green dashed curves are fit result for $J^{PC}=0^{-+}$, the blue dotted curves are $J^{PC}=1^{++}$, the red dash-dotted curves are $J^{PC}=2^{-+}$, and rose solid are $J^{PC}=2^{++}$.}\label{fig:fit_cos}
\end{figure*}

To further understand the nature of this enhancement, the helicity angles of the $\phi$ and $\Lambda$ candidates are studied.
The helicity angle is defined as the angle between the momentum of the $\phi$ or $\Lambda$ in its parent's rest frame and the momentum of $\phi$ or $\Lambda$'s parent in its grandparent's rest frame.
The helicity angular distributions for events in the $\phi$ signal region after efficiency correction are shown in Fig.~\ref{fig:fit_cos}, combining all data samples.
The unbinned maximum likelihood fit is performed simultaneously to the angular distributions, considering the same components as the ones contributing to the fit of the $M(\LLbar)$ distribution.
Fraction of each component is fixed to that obtained from the $\LLbar$ mass spectra fit.
Possible interference is not considered in the fit.
The shapes of resonant signal are described with the formula constructed according to Ref.~\cite{HelGen}.
The details of the formula we used are provided in the supplementary material.
The shapes of the background and PHSP signal distributions are assumed to be flat.
The number of events in each component is fixed to the fit result of $M(\LLbar)$.
The fit favors the hypotheses of $J^{PC}=1^{++}$, $2^{++}$, or $2^{-+}$, where the results lead to the similar fit quality.
The hypothesis of this enhancement having spin zero is rejected with significance greater than 7$\sigma$ compared with other hypotheses.
The fit result with different $J^{PC}$ hypotheses are shown in Fig.~\ref{fig:fit_cos}.

\begin{figure}[htbp]
\begin{center}
 \centering
 \includegraphics[width=0.45\textwidth]{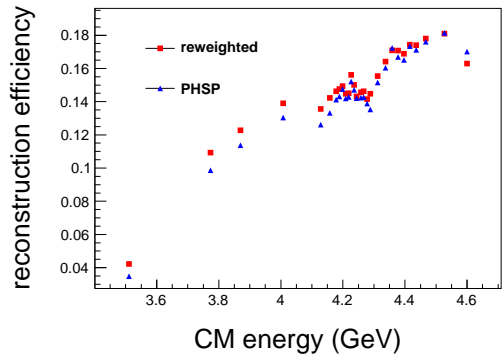}
 \caption{Energy-dependent reconstruction efficiencies using MC samples generated using a PHSP model (blue triangle) and after re-weighting the MC sample with the data distribution (red box).
 }
 \label{fig:eff}
 \end{center}
\end{figure}

Data driven reconstruction efficiencies are obtained by re-weighting the signal MC samples in the generator level.
The contribution of the near-$\LLbar$-threshold enhancement and non-uniform angular distributions measured in data are considered.
The energy-dependent reconstruction efficiencies for the PHSP and the re-weighted models are shown in Fig.~\ref{fig:eff}.
The fine structures observed in the efficiency curve is due to the deformation in the cross section line shape which is considered in MC generation to obtain the correct ISR factor and efficiency.

\subsection{Cross section measurement}

The cross section at a certain CM energy is calculated as
\begin{equation}
     \sigma= \frac{N_{\rm sig}}{\mathcal{L}_{\rm int} \mathcal{B} \varepsilon (1+\delta)},
\end{equation}
where $N_{\rm sig}$ is the number of $\phi\LLbar$ signal events obtained from the fit to the $M(K^{+}K^{-})$ distribution,
$\mathcal{L}_{\rm int}$ is the integrated luminosity,
$\varepsilon$ is a weighted value of the efficiencies from the process $\EE\to\phi\LLbar$ where $M(\LLbar)$ following the lineshape of the near-threshold enhancement as well as the angular distributions, and the process $\EE\to\phi\LLbar$ uniformly distributed in phase space, $\mathcal{B}$ is the product of the branching fraction of the intermediate decays $\phi\to\kk$ and $\Lambda\to p\pi$, which are taken from Ref.~\cite{PDG},
$(1+\delta)$ is the ISR correction factor. 

To obtain the proper ISR correction factor, an iterative procedure is used.
First, a series of signal MC samples are generated for all energy points with a constant cross section using {\sc kkmc}.
The cross sections are calculated based on the reconstruction efficiencies and ISR correction factors obtained from the signal MC simulation.
We use the Lowess~\cite{Lowess} method to smoothen the lineshape of the measured cross sections,
then use the method introduced in Ref.~\cite{wll_isr} to get the ISR correction factors and efficiencies with the new lineshape.
A new series of cross sections could be obtained, and after several iterations, the cross section results become stable.

\begin{figure}[htbp]
\begin{center}
 \centering
 \includegraphics[width=0.45\textwidth]{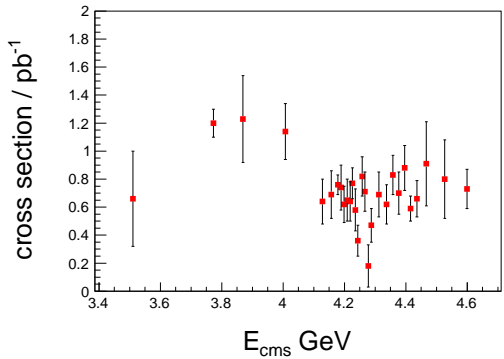}
 \caption{Energy-dependent cross section distribution, where dots with error bars correspond to the cross section measurements at each CM energy point.}
 \label{fig:BCS}
 \end{center}
\end{figure}

However, when the iteration is performed, the cross section results at each energy point are correlated.
To take the correlation into consideration, we use pseudoexperiments.
First, a large pseudodata sample is generated by sampling 
a Gaussian distribution, the mean value of which is the nominal cross section result, and its width is the statistical error from the fit to the data.
Then, the iteration described in the previous paragraph with this new lineshape is performed.
The resulting cross section distributions at each energy point are fitted with Gaussian functions.
Their mean and width values are taken as the final results for the cross sections and their corresponding uncertainties, respectively.
The final results are shown in Fig.~\ref{fig:BCS}, and listed in Table~\ref{tab:CS}, including the statistical and systematic uncertainties.

\section{Systematic Uncertainties}

\subsection{Uncertainties on Cross Sections}

The systematic uncertainties include contributions from luminosity, tracking and
PID efficiencies of the kaons, $\Lambda$ reconstruction, radiative correction factor associated with the efficiency, background, and branching fractions of the
intermediate states.

The integrated luminosity is measured using Bhabha scattering events, with uncertainty smaller than $1.0\%$~\cite{2015MAblikimLumi}.
The uncertainty related to the tracking efficiency of kaons is estimated to be 1.0\%, and the uncertainty arising
from the kaon PID efficiency is determined to be 1.0\% using a control sample $\EE\to\kk\pp$.
With the control sample, the tracking or PID requirement efficiency is separately measured in the MC simulation sample and in the data sample.
The difference between the efficiencies from MC simulation and data samples is taken as the systematic uncertainty.

The systematic uncertainty due to the $\Lambda$ reconstruction efficiency including the tracking and PID of its decay products $p\pi$, as well as the decay length requirement, is studied with the control sample $\Lambda_c\to\Lambda + X$ decays.
The resulting systematic uncertainty is 1.1\%~\cite{Lambda_c}.

The systematic uncertainty due to the $\Lambda(\bar{\Lambda})$ and $\phi$ mass window selection criteria accounts for the mass resolution discrepancy between the MC simulation and experimental data.
The $\phi$ and $\Lambda$ mass distributions from signal MC sample and data are fitted with double-Gaussian functions and compared with each other.
The difference between the fit results is negligible.

For the uncertainty due to the ISR correction factor, we change the lineshape with a power law function $1/s^{n}$.
The difference between the nominal result and the alternative parameterization is taken as the systematic uncertainty.
To estimate the systematic uncertainty related to the background model, we vary the fit range of the $M(\kk)$ distribution. We also use a second-order polynomial as an alternative background model. The largest value among all variations is taken as systematic uncertainty for this source.
Due to limited sample sizes at most CM energy points, the uncertainty from the data sample collected at $\sqrt{s}=4.178~\rm GeV$ is used for all the data sets.

The uncertainty of the kinematic fit is estimated by comparing the reconstruction efficiency before and after the helix parameter correction using the method described in Ref.~\cite{helix}.
It should be pointed out that the reconstruction efficiency after the helix parameter correction is used as the nominal result.
The uncertainties of the branching fractions are taken from the PDG~\cite{PDG}.

The summary of the systematic uncertainties at $\sqrt{s}=4.178~\rm GeV$ is presented in Table~\ref{tab:SysErr}.

\begin{table}[!htb]
  \centering
  \caption{Summary of the systematic uncertainties for the cross section measurements using $\sqrt{s}=4.178~\rm GeV$ data as an example.}
  \label{tab:SysErr}
  \begin{tabular}{cc}
  \hline\hline
  Source  &  (\%) \\ \hline
  Luminosity  &  1.0 \\
  Tracking  &  2.0 \\
  PID  & 2.0 \\
  $\Lambda$ reconstruction  &   1.1  \\
  ISR factor  &  5.6 \\
  Kinematic fit  &  1.0 \\
  Branching fraction  &  0.8 \\
  Background model &  2.6 \\ \hline
 Total  & 7.1  \\
  \hline
  \hline
  \end{tabular}
\end{table}

\subsection{Uncertainties on $M(\LLbar)$ Lineshape}

The systematic uncertainties for the mass and width of the BW formed lineshape include those from the mass calibration, efficiency curve, signal parameterization, and background estimation.

To calibrate the mass component, a maximum likelihood fit of the $K^{+}K^{-}$ invariant mass distributions is performed for all the data samples.
The difference between the fitted mass and the known mass of the $\phi$ meson~\cite{PDG} is $0.4~{\rm MeV}/c^{2}$.
According to the conservation of energy and momentum, the difference on the $\LLbar$ side should be 0.3~${\rm MeV}/c^2$. This value is taken as the systematic uncertainty.

To evaluate the systematic uncertainty from the efficiency curve estimation, we use unweighted PHSP MC sample instead of the nominal one to extract the efficiency curve.
The changes on the mass and width, 0.6 MeV/$c^2$ and 8.0 MeV, respectively, are taken as the systematic uncertainties.

To account for the systematic uncertainty from the signal model, we change the parameterization form from a Landau to a BW function.
The differences between the two parameterizations, 22.9 ${\rm MeV}/c^2$ and 13.5 $\rm MeV$, are taken as the systematic uncertainties on the mass and width, respectively.
Another source of uncertainty in the signal parameterization is the quantum number assumption. We change the assignment of $l/L_d=0/1$ to $1/2$.
The differences on the mass and width, $1.7~{\rm MeV}/c^2$ and $36.0~\rm MeV$, are taken as the systematic uncertainties.
For the background estimation, we get different yields by varying the fit range of the $K^{+}K^{-}$ invariant mass distributions and repeating the fit.
The differences on the mass and width, 3.1 ${\rm MeV}/c^{2}$ and 1.2 $\rm{MeV}$, respectively, are taken as the systematic uncertainties.
Replacing the background parameterization with a BW function leads to changes in the measurement of the mass and width, $15.9~{\rm MeV}/c^2$ and $18.0~\rm MeV$, which are are taken as the systematic uncertainties from the background model.

Table \ref{tab:uncertainties2} summarizes these three sources of uncertainties. The sum of all the above uncertainties in quadrature,  28.1~${\rm MeV}/c^{2}$ and 43.2~MeV for the mass and width, respectively, are taken as the total uncertainties.

The systematic uncertainty for the reversed exponential parameter come from similar sources.
We broaden or narrow the fit range by 0.02 GeV, and the change of the exponential parameter 4.8 is taken as systematic uncertainty.
Efficiency curve is flat near the threshold, which will not affect the exponential parameter.
The mass calibration will not change the lineshape so this source is also ignored.
We vary the background estimation and background shape, the largest changes of the exponential parameter is 0.4.
The uncertainties are also summarized in Table~\ref{tab:uncertainties2}.
The sum of all the above uncertainties in quadrature, $4.8$, is taken as the total systematic uncertainty.

\begin{table}[!htb]
  \centering
  \caption{Summary of systematic uncertainties for the $M(\LLbar)$ near-threshold enhancement lineshape parameters. }
  \label{tab:uncertainties2}
  \begin{tabular}{cccc}
  \hline\hline
  Source  &  Mass (${\rm MeV}/c^2$)  &   Width (MeV) & $p_{0}$ (MeV/$c^2$)  \\ \hline
  Mass calibration  &   0.3   &   $-$ & $-$ \\
  Efficiency curve  &   0.6   &   8.0 & $-$ \\
  Signal model   &   22.9   &   13.5 & $-$   \\
  Quantum number &  1.7	&  36.0 & $-$	\\
  Background estimation   &  3.1   &    1.2 & $0.4$   \\
  Background model   &  15.9   &    18.0  & $-$  \\
  Fit range & $-$ & $-$ & $5.5$ \\
  Total   &    28.1  &   43.2  & $5.5$  \\
  \hline\hline
  \end{tabular}
\end{table}

\section{Summary and Discussion}

In summary, we observe the $\EE \to \phi \Lambda \bar{\Lambda}$
process for the first time with data samples at CM energies ranging
from 3.51 to 4.60~GeV. The energy-dependent cross sections of
$\EE \to \phi \Lambda \bar{\Lambda}$ are measured. Due to the
limited sample sizes, we cannot resolve the composition of the
resonance structure, and the lineshape might not be
simply described with a continuum process parameterized as $1/s^n~(n=3.3\pm0.3)$.

Moreover, a near-threshold enhancement is observed on 
$\Lambda\bar{\Lambda}$ with a significance greater
than $25\sigma$ compared with the pure phase space distribution. 
By fitting the lineshape with a BW function, we obtain
the mass and width as $(2262 \pm 4 \pm 28)~{\rm MeV}/c^2$ and
$(72 \pm 5 \pm 43)~\rm MeV$, repectively, where the first uncertainties are
statistical and the second ones systematic. 
By fitting the linehshape with an reversed exponential function, we obtain the rising rate 
(exponential parameter) as $33\pm11\pm6~\rm MeV/c^2$.

According to the helicity angle study, the quantum numbers of $\LLbar$ system
$J^{PC}=0^{++/-+}$ is rejected with a significance of $7\sigma$. The
interpretation of the $\LLbar$ system originating from a decay
$\eta(2225)\to\LLbar$ is rejected. The $J^{PC}$ quantum numbers
could be $2^{++}$, $2^{-+}$, or $1^{++}$, but they cannot be
distinguished because of the limited data sample sizes. 
Another interpretation of a lower mass resonance is that this could be a $f_2(2300)$ meson. 
However, according to previous measurements using the decay
modes $\jpsi\to \gamma\phi\phi$ and $\jpsi\to
\gamma\LLbar$~\cite{etaBESIII,J2rLL_BESIII}, the $f_2(2300)$ meson is
more likely to decay into a $\phi\phi$ final state rather than to
$\LLbar$. The cross sections of the
$\EE\to\phi\phi\phi$ process are measured~\cite{Ablikim:2017rnw} at BESIII
with similar cross sections to those of the $\EE\to\phi\LLbar$ process, but no
structure around $2.23~{\rm GeV}/c^2$ is observed in the
$\phi\phi$ mass spectrum. Therefore the interpretation of
$f_2(2300)\to\LLbar$ is also rejected.

This enhancement does not match any known resonance~\cite{PDG}
seen before, and could be the same thing observed in $B\to
K\LLbar$ decays~\cite{B2KLLbar}. If so, the theoretical
explanation, an isoscalar state with $J^{PC}=0^{\pm+}$ coupled to
a pair of gluons~\cite{Rosner:2003bm}, could be discarded since the
hypothesis implying quantum numbers $0^{-+}$ is rejected based on
the angular distribution study. Also, the author of
Ref.~\cite{Rosner:2003bm} implies that the observed threshold
enhancements in low-mass baryon-antibaryon systems might not be
limited to the ordinary quantum numbers of the $q\bar{q}$ system.
Further studies of the $\LLbar$ system would be helpful to
understand the nature of this threshold enhancement. For example,
a search for a threshold enhancement in the $\EE\to \eta\LLbar$ process could provide a crucial test because
the states produced in this mode have the exact opposite {\it C} parity to the state in our analysis.

\section{Acknowledgements}
The BESIII collaboration thanks the staff of BEPCII and the IHEP computing center for their strong support. This work is supported in part by National Key Research and Development Program of China under Contracts Nos. 2020YFA0406300, 2020YFA0406400; National Natural Science Foundation of China (NSFC) under Contracts Nos. 11625523, 11635010, 11735014, 11822506, 11835012, 11935015, 11935016, 11935018, 11961141012; the Chinese Academy of Sciences (CAS) Large-Scale Scientific Facility Program; Joint Large-Scale Scientific Facility Funds of the NSFC and CAS under Contracts Nos. U1732263, U1832207; CAS Key Research Program of Frontier Sciences under Contracts Nos. QYZDJ-SSW-SLH003, QYZDJ-SSW-SLH040; 100 Talents Program of CAS; INPAC and Shanghai Key Laboratory for Particle Physics and Cosmology; ERC under Contract No. 758462; European Union Horizon 2020 research and innovation programme under Contract No. Marie Sklodowska-Curie grant agreement No 894790; German Research Foundation DFG under Contracts Nos. 443159800, Collaborative Research Center CRC 1044, FOR 2359, FOR 2359, GRK 214; Istituto Nazionale di Fisica Nucleare, Italy; Ministry of Development of Turkey under Contract No. DPT2006K-120470; National Science and Technology fund; Olle Engkvist Foundation under Contract No. 200-0605; STFC (United Kingdom); The Knut and Alice Wallenberg Foundation (Sweden) under Contract No. 2016.0157; The Royal Society, UK under Contracts Nos. DH140054, DH160214; The Swedish Research Council; U. S. Department of Energy under Contracts Nos. DE-FG02-05ER41374, DE-SC-0012069.

\newpage
\appendix
\part*{Supplemental material}
\addcontentsline{toc}{part}{Appendices}

\section{Simultaneous fit of $M(\LLbar)$}

The center-of-mass energy dependent simultaneous fit result of $M(\LLbar)$ is are shown in Fig.~\ref{fig:SimFit}

\begin{figure*}[htbp]
\begin{center}
 \centering
	 \includegraphics[width=0.99\textwidth]{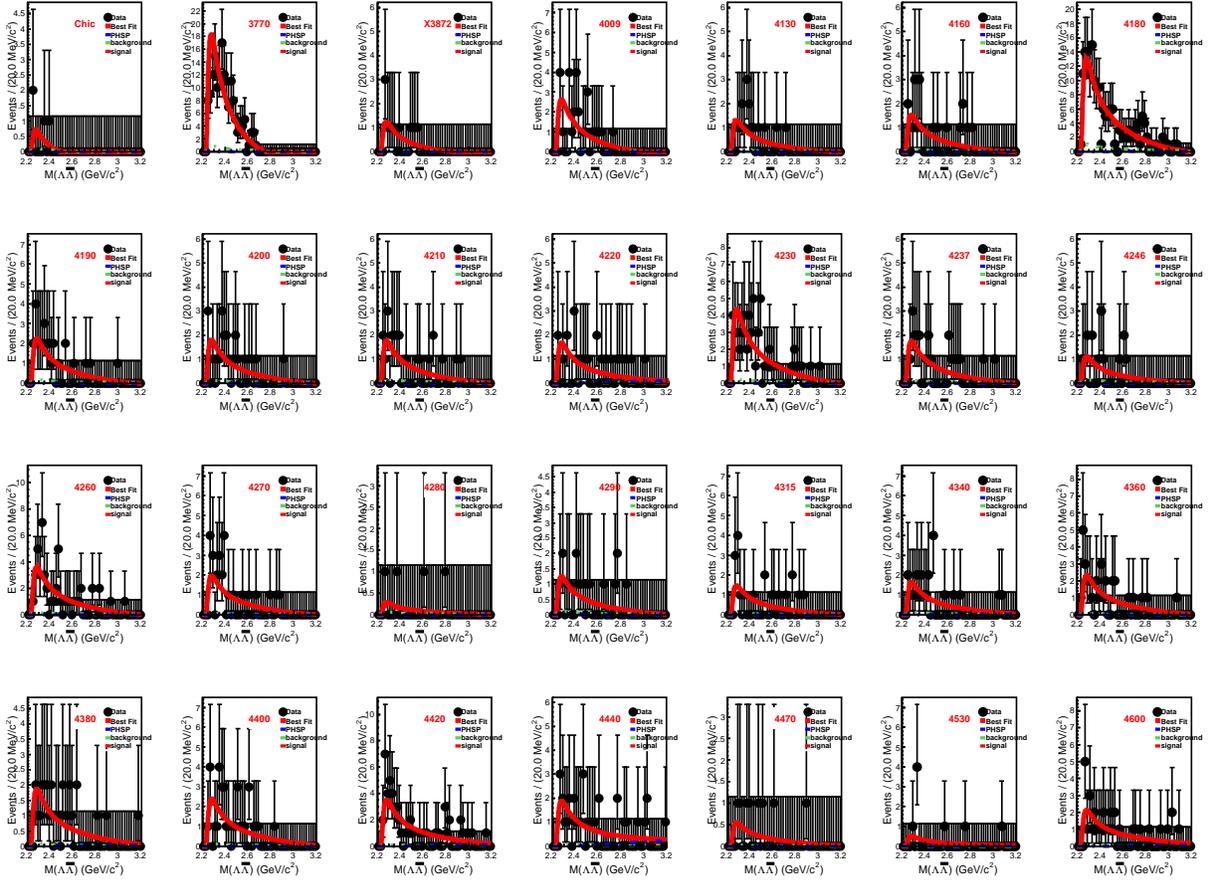}
 \caption{Simultaneous fit to $\rm{M}(\Lambda\bar{\Lambda})$ for all data samples.}
 \label{fig:SimFit}
 \end{center}
\end{figure*}

\section{Helicity Angular Distribution Evalution}
The angular distribution parameterization is constructed according to Ref.~\cite{HelGen}.
The differential cross section
\begin{equation}
\begin{aligned}
\frac{d\sigma}{d\Omega} = \sum_{M,\mu,\mu',\nu'}|\sum_{\nu}A(M)|^2,
\end{aligned}
\end{equation}
where $M=\pm 1$ is the magnetic moment of $e^+e^-$ system, $\mu,\nu$ are the helicity eigenstates of $\phi$, $X(2260)$, $\mu',\nu'$ are the helicity eigenstates of $\Lambda$ and $\bar{\Lambda}$,
and 
\begin{equation}
\begin{aligned}
A(M) = C_1F^{i}_{\mu,\nu}D^{i}_{M,\mu-\nu}F^{j}_{\mu',\nu'}D^{j}_{\nu,\mu'-\nu'},
\end{aligned}
\end{equation}
where a $C_1$ is constant parameter which we ignore, $F^{i}_{\mu,nu}$ and $F^{j}_{\mu',\nu'}$ are the amplitudes where $i=1$ is the spin of $e^+e^-$
system, and $j$ is the spin of $\LLbar$ system, and $D^{i}_{M,\mu-\nu}$ is the Wigner D-matrix.
After substituting the different hypothesis for $J^{PC}$ in the Wigner D-matrix, the formula is simplified and integrated over the the azimuthal angle, we obtain the angular distribution of the final state for different hypotheses.

\begin{figure*}[htbp]
\begin{center}
    \includegraphics[width=0.99\textwidth]{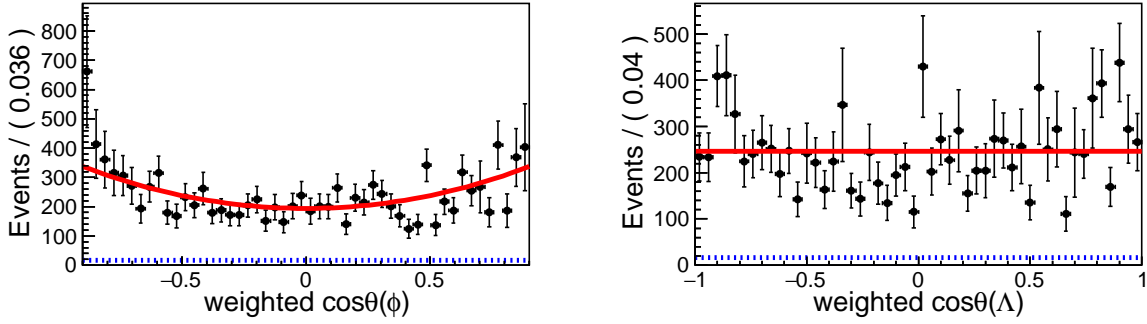}
\end{center}
\caption{Efficiency-corrected angular distribution of the $\phi$ (left) and $\Lambda$ (right) candidates combining all the data samples and the simultaneous fit results with the hypotheses of $J^{PC}=0^{-+}$. The data (dots) are overlaid by the result of the fit (red solid line) described in the text, the blue dotted curve are the components consisting of the PHSP signal and the sideband background.}\label{fig:fit_cos_0mp}
\end{figure*}

For the hypotheses of $0^{-+}$ and $0^{++}$ cases, the Wigner D-matrix $D^{0}_{M,\mu-\nu}$ is a constant, which means the helicity angle distribution of $\Lambda(\bar{\Lambda})$ should be uniform.
The fitting results with $J^{PC}=0^{-+}$ hypothesis is shown in Fig.~\ref{fig:fit_cos_0mp} as an example.
Clearly the fitting result on $\Lambda(\bar{\Lambda})$ helicity angular distribution is bad.

For $J^{PC}=1^{++}$, we have

\begin{equation}
\begin{aligned}
\frac{d\sigma}{d{\rm cos}\theta_{\phi}} \propto A+B{\rm cos}^2\theta_{\phi} \\
\frac{d\sigma}{d{\rm cos}\theta_{\Lambda}} \propto a+b{\rm cos}^2\theta_{\Lambda}, 
\end{aligned}
\end{equation}

where

\begin{equation}
\begin{aligned}
A = F_{01}+F_{10}+2F_{11} 
\end{aligned}
\end{equation}
\begin{equation}
\begin{aligned}
B = F_{01}+F_{10}-2F_{11} 
\end{aligned}
\end{equation}
\begin{equation}
\begin{aligned}
a = F_{01}+F_{11} 
\end{aligned}
\end{equation}
\begin{equation}
\begin{aligned}
b = 2F_{10}-F_{01}-F_{11} ,
\end{aligned}
\end{equation}

\begin{figure*}[htbp]
\begin{center}
    \includegraphics[width=0.99\textwidth]{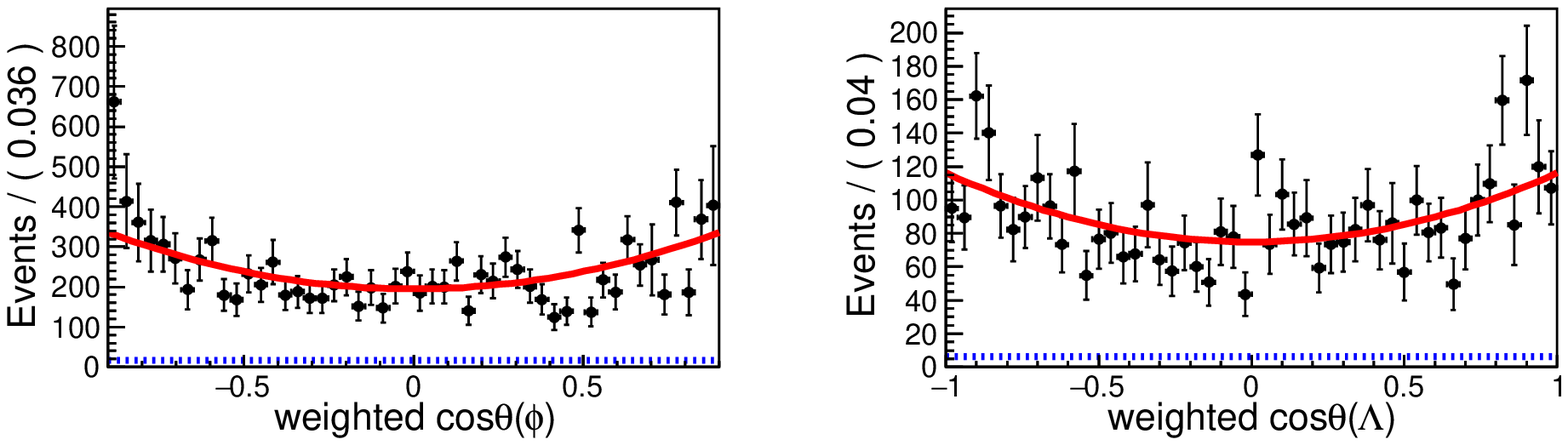}
\end{center}
\caption{Efficiency-corrected angular distribution of the $\phi$ (left) and $\Lambda$ (right) candidates combining all the data samples and the simultaneous fit results with the hypotheses of $J^{PC}=1^{++}$. The data (dots) are overlaid by the result of the fit (red solid line) described in the text, the blue dotted curve are the components consisting of the PHSP signal and the sideband background.}\label{fig:fit_cos_1pp}
\end{figure*}

The fitting results with $J^{PC}=1^{++}$ hypothesis is shown in Fig.~\ref{fig:fit_cos_1pp}.

For $J^{PC}=2^{++}$, we have

\begin{equation}
\begin{aligned}
	\frac{d\sigma}{d{\rm cos}\theta_{\phi}} \propto A+B{\rm cos}^2\theta_{\phi} \\
	\frac{d\sigma}{d{\rm cos}\theta_{\Lambda}} \propto a+b{\rm cos}^2\theta_{\Lambda}+c{\rm cos}^4\theta_{\Lambda}, 
\end{aligned}
\end{equation}

where

\begin{equation}
\begin{aligned}
	A = F_{00}+F_{01}+F_{10}+2F_{11}+F_{12}\\
\end{aligned}
\end{equation}
\begin{equation}
\begin{aligned}
	B = F_{01}+F_{10}-F_{00}-2F_{11}+F_{12}\\
\end{aligned}
\end{equation}

\begin{equation}
\begin{aligned}
	a = F_{01}F_{-+}+F_{11}F_{-+}+F_{12}F_{-+}+\frac{1}{2}F_{00}F_{++}+F_{10}F_{++}+\frac{3}{2}F_{12}F_{++} \\
\end{aligned}
\end{equation}

\begin{equation}
\begin{aligned}
	b = 3(F_{00}F_{-+}-F_{01}F_{-+}+2F_{10}F_{-+}-F_{11}F_{-+}\\
		-F_{00}F_{++}+2F_{01}F_{++}-2F_{10}F_{++}+2F_{11}F_{++}-4F_{12}F_{++}) 
\end{aligned}
\end{equation}

\begin{equation}
\begin{aligned}
	c = -3F_{00}F_{-+}+4F_{01}F_{-+}-6F_{10}F_{-+}+4F_{11}F_{-+}-F_{12}F_{-+}+ \\
		\frac{15}{2}F_{00}F_{++}-6F_{01}F_{++}+9F_{10}F_{++}-6F_{11}F_{++}+\frac{3}{2}F_{12}F_{++} 
\end{aligned}
\end{equation}

\begin{figure*}[htbp]
\begin{center}
    \includegraphics[width=0.99\textwidth]{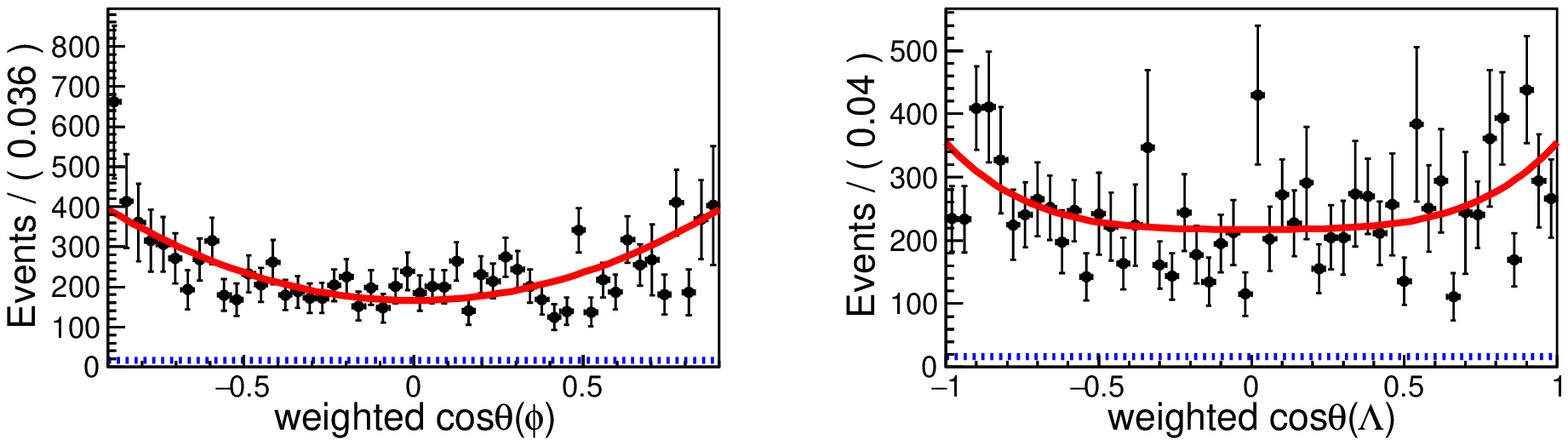}
\end{center}
\caption{Efficiency-corrected angular distribution of the $\phi$ (left) and $\Lambda$ (right) candidates combining all the data samples and the simultaneous fit results with the hypotheses of $J^{PC}=2^{++}$. The data (dots) are overlaid by the result of the fit (red solid line) described in the text, the blue dotted curve are the components consisting of the PHSP signal and the sideband background.}\label{fig:fit_cos_2pp}
\end{figure*}

The fitting results with $J^{PC}=2^{++}$ hypothesis is shown in Fig.~\ref{fig:fit_cos_2pp}.

For $J^{PC}=2^{-+}$, we have
\begin{equation}
\begin{aligned}
	\frac{d\sigma}{d{\rm cos}\theta_{\phi}} \propto A+B{\rm cos}^2\theta_{\phi} 
\end{aligned}
\end{equation}
\begin{equation}
\begin{aligned}
	\frac{d\sigma}{d{\rm cos}\theta_{\Lambda}} \propto a+b{\rm cos}^2\theta_{\Lambda}+c{\rm cos}^4\theta_{\Lambda},
\end{aligned}
\end{equation}

where

\begin{equation}
\begin{aligned}
A = F_{01}+F_{10}+2F_{11}+F_{12}\\
\end{aligned}
\end{equation}
\begin{equation}
\begin{aligned}
B = F_{01}+F_{10}-2F_{11}+F_{12}\\
\end{aligned}
\end{equation}
\begin{equation}
\begin{aligned}
a = 2F_{10}+3F_{12} \\
\end{aligned}
\end{equation}
\begin{equation}
\begin{aligned}
b = 12F_{01}-12F_{10}+12F_{11}-6F_{12} \\
\end{aligned}
\end{equation}
\begin{equation}
\begin{aligned}
c = -12F_{01}+18F_{10}-12F_{11}+3F_{12} 
\end{aligned}
\end{equation}

\begin{figure*}[htbp]
\begin{center}
    \includegraphics[width=0.99\textwidth]{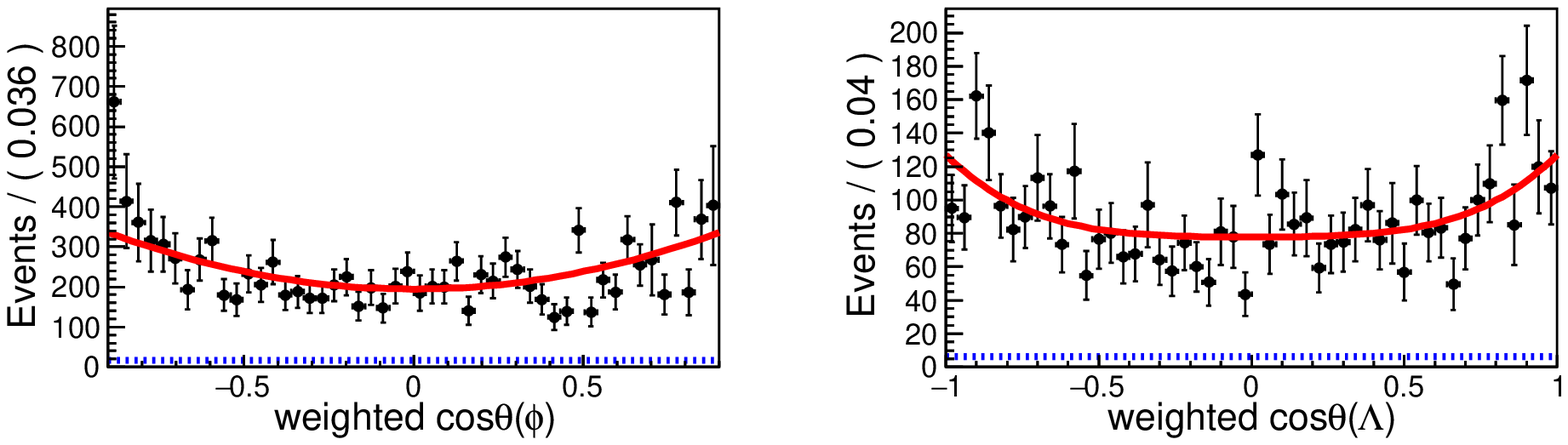}
\end{center}
\caption{Efficiency-corrected angular distribution of the $\phi$ (left) and $\Lambda$ (right) candidates combining all the data samples and the simultaneous fit results with the hypotheses of $J^{PC}=2^{-+}$. The data (dots) are overlaid by the result of the fit (red solid line) described in the text, the blue dotted curve are the components consisting of the PHSP signal and the sideband background.}\label{fig:fit_cos_2mp}
\end{figure*}

The fitting results with $J^{PC}=2^{-+}$ hypothesis is shown in Fig.~\ref{fig:fit_cos_2mp}.

\end{document}